**RSCPublishing Soft Matter**

## Defect Textures in Polygonal Arrangements of Cylindrical Inclusions in Cholesteric Liquid Crystal Matrices

| | |
|---:|:---|
| Journal: | *Soft Matter* |
| Manuscript ID: | Draft |
| Article Type: | Paper |
| Date Submitted by the Author: | n/a |
| Complete List of Authors: | Murugesan, Yogesh Kumar; McGill University, Department of Chemical Engineering<br>Pasini, Damiano; McGill University, Department of Mechanical Engineering<br>Rey, Alejandro; McGill University, Department of Chemical Engineering |

**SCHOLARONE™ Manuscripts**

# Defect Textures in Polygonal Arrangements of Cylindrical Inclusions in Cholesteric Liquid Crystal Matrices

Yogesh K. Murugesan[a], Damiano Pasini[b]  Alejandro D. Rey[a*]

[a]Department of Chemical Engineering, McGill University, 3610 University Street, Montreal, Quebec, H3A 2B2, Canada. E-mail: alejandro.rey@mcgill.ca; Fax: +1 514 398-6678; Tel: +1 514 398-4196
[b]Department of Mechanical Engineering, McGill University, 817 Sherbrooke West, Montreal, Quebec, H3A 2K6, Canada
 *corresponding author

**Abstract**

A systematic computational and scaling analysis of defect textures in polygonal arrangement of cylindrical particles embedded in a cholesteric (Ch) liquid crystal matrix is performed using the Landau de Gennes model for chiral self assembly, with strong anchoring at the particles' surface. The defect textures and LC phases observed are investigated as a function of chirality, elastic anisotropy (monomeric and polymeric mesogens), particle polygonal arrangement and particle size. The presence of a polygonal network made of N circular inclusions results in defect textures of a net charge of $-(N-2)/2$ per unit polygonal cell, in accordance with Zimmer's rule. As the chirality increases, the LC matrix shows the following transition sequence: weakly twisted cholesterics, 2D blue phases with non-singular/ singular defect lattices, cholesteric phases with only disclinations, and finally fingerprint cholesteric textures with disclinations and dislocations. For monomeric mesogens at concentrations far from the I/Ch phase transition and low chirality, for a given symmetry of the LC phase, the particle with weaker (stronger) confinement results in a phase with lower (higher) elastic energy, while at high chirality the elastic energy of a LC phase is proportional to the number of particles that form the polygonal network. Thus, hexagonal (triangular) particle arrangement results in low elastic energy at low (high) chirality. For semiflexible polymeric mesogens (high elastic anisotropy), defect textures with fewer disclinations/ dislocations arise but due to layer distortions we find a higher elastic energy than monomeric mesogens. The defects arising in the simulations and the texture rules established are in agreement with experimental observations in cellulosic liquid crystal analogues such as plant cell wall and helical biological polymeric mesophases made of DNA, PBLG and xanthan. A semi-quantitative phase diagram that shows different LC phase and defect textures as a function of chirality and elastic anisotropy is obtained. The inclusion of particles has a stabilizing effect on the LC phases, as they occupy $\lambda^{+1}$ disclination cores, thereby reducing the free energy cost associated with these disclinations. The findings provide a comprehensive set of trends and mechanisms that contribute to the evolving understanding of biological plywoods and serve as a platform for future biomimetic applications.

# 1 Introduction

Particles in liquid crystalline (LC) mesophases are of increasing interest from both fundamental physics and engineering applications point of view. Biological fibrous composites[1,2], carbon-fiber-reinforced carbon composites[3], water in cholesteric (Ch) LC emulsion[4], polymer dispersed liquid crystals[5] and filled nematics[6], cholesterics[6,7] and smectics[8,9] are some examples of biphasic heterogeneous materials consisting of suspensions of secondary phases such as colloidal particles, nanoparticles, drops, or aligned fibers in a LC matrix. These materials can spontaneously arise through phase separation processes between a polymer melt and a LC material, through dispersion of the secondary phase[1,2], or through flow of LC mesophase via (too many through here is confusing) aligned fibers[3]. The former occurs in synthesis of biological fibrous composites such as plant cell walls[1] and insect cuticles[2] in nature, where biopolymers in extracellular matrix self-assemble to form cholesteric (cholesteric) mesophases in the presence of secondary phases such as cell lumen and pit canals in plant cell walls, and pore canals in insects. The latter forms the basis of carbon/carbon high performance composites obtained by injecting a carbonaceous discotic nematic mesophase into the fiber bundle[3]. Since liquid crystals possess long range order and gradient orientation elasticity[10], under sufficiently strong interaction between the fibrous (cylindrical) inclusions and the LC mesophase, the direction of orientational anchoring at the fiber's curved interface propagates through the gradient elasticity effect into the bulk. This results in texturing through nucleation of bulk orientational defects, such as disclinations and dislocations and through defect coarsening and defect splitting. Computational and experimental investigation of the nature, density and positional order of defects associated with embedded fibers and with fiber assemblies in filled LC phases such as nematic, chiral-nematic, smectic, columnar are the at the core of an evolving field. The key motivation of this paper is to investigate the defect texturing mechanism in cholesteric mesophases with cylindrical inclusions, which is relevant to biological and synthetic systems, and serves as a biomimetic system model for functional and structural materials.

The cylindrical particle-filled cholesteric mesophases studied in this work are known to arise in biological fibrous composites such as plant cell walls[1] and insect cuticles[2] in nature. These biological materials are termed as LC analogues, where the LC type of packing is observed frozen-in in the solid state[11]. In plant cell walls, cellulose microfibrils coated with hemicelluloses and embedded in a lignin/pectin matrix[12] are known to self assemble, and form structures called the twisted plywood architecture[11] and defect patterns characteristics of cholesteric mesophases[2,11]. Recently, a mathematical model based on the Landau-de Gennes theory of liquid crystals has been used to simulate defect textures arising in some plant cell walls substantiate the role of LC self-assembly in the formation of twisted plywood architecture in plant cell walls[13]. The domain of this supramolecular self assembly is rich in secondary inclusions including: (a) pit canals[2] of diameter of about 1 μm, regularly spaced and evenly distributed over the cell surface (in the range of 5 to 7 per μm$^2$) resembling dispersed hollow cylinders, and (b) elongated plant cells[2] which produce the cell wall components. These micron-sized cylindrical inclusions are known to form the constraining surface that can direct the growth of a cholesteric helix in a direction normal to it, necessary to form a defect free monodomain twisted plywood structure[11,14]. A significant body of experimental literature indicates that the polygonal network as shown in Fig. 1, obtained by joining the centers of secondary inclusions in some plant cell walls, includes triangles, rectangles, pentagons, and hexagons[2].

Similar polygonal network of particle assemblies embedded in LC mesophase are known to arise in synthetic composites such as random heterogeneous filled nematics[3,15], crystalline filled nematics[5] and micellar cholesterics[16]. The defect texturing mechanism in particle embedded nematic phases in these synthetic composite systems has been investigated using theory and simulation[17]. The defect texture selection in a nematic carbonaceous mesophase matrix with embedded circular carbon fibers is known to satisfy Poincaré-Brouwer's theorem; the net disclination line strength is equal to $-(N-2)/2$, where N is the number of inclusions[17]. For submicron sized fibers, the number of defects is equal to $(N-2)$ and all the disclinations are singular, such that the nanoscale core has a biaxial molecular order with a negative uniaxial center; defect types are discussed in more detail below. For micron-sized cylindrical fibers, the defect texture is dependent on fiber configuration i.e., for an even number of particles there is a single defect with an escaped micron-sized core and for odd number of particles, the $(N-2)$ disclinations nucleated have singular nano-scale cores[17]. A computational model developed for coupled phase separation-phase ordering processes has been used to resolve defect nucleation, defect-defect interactions and defect-particle interactions in polymer/liquid crystal mixtures[5]. When numerically solved, the model can predict the emergence of a defect lattice with fourfold symmetry in droplet phase-separated morphologies under spinodal decomposition. The resulting nematic-droplet structure has a stable polymer droplet crystal lattice pinned by a lattice of topological defects and the charge associated with each embedded polymer drop has been computed to be +1[5]. Recently[15], a series of computational investigations on texture formation mechanism in polygonal arrangement of faceted particles embedded in a nematic LC matrix concluded that the excess free energy per particle is dependent on the embedded particle configuration. Excess energy per particle for even sided polygonal networks (rectangular and hexagonal) have been reported to be lower than that of odd sided particle networks (triangular and pentagonal)[15]. Despite common occurrence of polygonal networks in cholesteric LCs in many biological fibrous composites[1,2,11] and micellar cholesterics with embedded infinite cylindrical micelles[16], the knowledge of defect nucleation and texturing rules in these chiral systems is far from complete. Cholesteric LCs with cylindrical inclusions being lamellar soft matter with curved interfaces, the defect texturing in these systems involves a delicate interplay between symmetry, number and geometry of embedded particle arrangement, confinement of the LC phase, viscoelastic properties of the LC material and multiple length scales arising due to diverse geometric features such as particles, pitch, domain size, interfaces, and defect cores. Understanding defect texturing is crucial in developing potential applications such as biomimetic material processing[18], supramolecular templating[8], spatial distribution control of nanoparticles[19], stabilization of LC phases[4,20]. Recently[13], the Landau de Gennes continuum model with a **Q** tensor representation of the liquid crystal orientation has been used to simulate defect textures arising in the domain of self assembly, due to presence of a rectangular network formed by four cell lumens embedded in the cell wall of a walnut specimen. The number and type of defects arising in all the simulations obeys the rule that the net charge of defects nucleated in a domain confined by four circular inclusions is always −1. The total number of defects nucleated is inversely proportional to the penetration length, which is the dilation strain required to produce bending in a set of parallel layers and it is smaller than the cholesteric pitch[13]. The main objective of this paper is to extend this model to triangular and hexagonal polygonal networks in an attempt to establish generic texture rules in cholesteric mesophases under spatial frustrations induced by the cylindrical inclusions with diverse symmetries. For an extensive review of mesoscopic models employed to quantitatively describe the biological liquid crystalline phases and processes see [21].

The cholesteric mesophase investigated in this study is a lamellar soft matter material that possesses long range orientational order with a superimposed rotation and layer periodicity[10]. The layer distance between a $2\pi$ rotation of the mesogens is the equilibrium pitch $p_0$. This is a periodic layered structure with layers of thickness half a helical pitch each. On a larger length scale, these systems have layered structure like a smectic liquid crystal and inside each layer the orientation is like that of a nematic liquid crystal. Owing to this dual structural nature, these phases nucleate orientational defects (disclinations) typical of nematic phases and translational defects (dislocations) typical of smectic phases. The possible mechanisms for defect nucleation in cholesteric mesophases with embedded cylindrical inclusions are as follows. (1) Spatial frustration (single particle mechanism)[22]: during chiral self assembly, the presence of cylindrical inclusions directs the cholesteric helix in more than one spatial direction, owing to the curvature of the particle- mesophase interface, resulting in spatial frustrations. These frustrations are relieved through the formation of blue phases that are composed of double twist cylinders and defect networks. The blue phases locally minimize the bulk free energy as opposed to defect free monodomain that minimizes the bulk free energy globally in a frustration-free chiral self-assembly[22] (2) Kibble mechanism[23] (particle-particle interaction): a large number of defects nucleate during the isotropic-cholesteric phase transition in presence of inclusions, when layers growing from adjacent particles impinge; defects arise due to the orientation incompatibilities between the coalescing cholesteric domains[23]. (3) Anisotropic confinement (multiparticle interactions): in mesophases, inclusion of secondary phases under strong anchoring induces confinement that leads to stress concentration around the particles that are relieved through nucleation of defects[21]. In presence of particle networks, the self-assembling material is subjected to spatially varying confinement owing to their particle configuration. Cholesteric mesophases respond to this frustration induced by confinement through bending and compression of layers and/or nucleation of disclinations and dislocations.

The texture $\tau^N$ in a cholesteric mesophase confined within a polygon of N sides obtained by joining the centers of N cylindrical inclusions of radius R, can be characterized by the number of defects (n), the type of defect core ($D_c$), and the total charge of the defects ($C_h$). The defect core type can be singular or non-singular (also known as escaped core); in the former the defect core is characterized by a complex radial nanoscale gradient in the molecular order parameters ($\tau$ lines), while in the latter the director escapes into the third dimension forming a macroscopic core region ($\lambda$ lines). The coreless non-singular disclination lines of core size of the order of $p_0$ have lower energy than the singular $\tau$ disclination lines of core size of the order of nanometers. Charge of an individual disclination can be $\pm 1/2$ or $\pm 1$ and describes the amount of director rotation when encircling the defect core. The dislocation core can be compact non-split core or split into a pair of disclinations. Figure 2 shows director configuration of some of the classical disclinations and disclination pairs arising in our simulations. Reference [24] presents a more extensive classification of defects in cholesteric LCs. The texture $\tau^N$ is a function of the ratio of distance between the inclusions measured along the line connecting the centers of two particles to the cholesteric pitch ($r_0/p_0$), the material parameter that represents a measure of the elastic anisotropy of the material ($\upsilon$), the thermodynamic potential proportional to the dimensionless concentration of the mesogens which drives the isotropic- cholesteric phase transition (U), the size of the cylindrical inclusion (R) and the number of inclusions (N) which in turn represents the confinement geometry and confinement intensity.

$$\tau^N(n, D_c, C_h) = \psi(p_0, \upsilon, R, U, N) \qquad (1)$$

In this paper, 2D defect textures in cholesteric mesophases phases with embedded networks of parallel cylindrical inclusions are investigated as a function of the length scale ratio $\xi/p_0$, the material property $\upsilon$, inclusion arrangement N, and particle size R, in a regime known as strong anchoring, where the orientation and molecular order at the particle surface are fixed and constant.

The specific objectives of this paper are: (a) to obtain defect texture rules in cholesteric mesophases with embedded polygonal particle networks in the ideal case of arrangements of particles with perfect order, (b) to characterize the defect texturing mechanism as a function of chirality, elastic anisotropy and particle configuration, and to establish a generic phase diagram, (c) to understand the interaction between particles and defect lattices and the role of particles on LC phase stability, (d) to identify the polygonal arrangement that may lower the elastic energy.

The organization of the paper is as follows. In Section 2, the quadrupolar order parameter that describes the orientation and degree of alignment of mesogens is introduced; the dimensionless form of Landau- de Gennes model for cholesterics, in terms of length scales of the model and material properties, is presented; and the dimensionless coarse-grained Lubensky- de Gennes model used to analyze some aspects of the results in terms of layer properties of the liquid crystalline phase, is also presented. In Section 3, the computational domain under investigation is introduced and the anisotropic confinement in the domain that results due to the chosen particle arrangement geometry is discussed. The imposed initial and boundary conditions, and the material properties used in the simulations are also presented. In Section 4, the defect textures arising in our simulation while exploring the viscoelastic-geometric parametric space are presented and the defects are characterized. The effect of chirality, elastic anisotropy, and inclusion configuration is investigated to understand the role of confinement, symmetry and layer deformation elasticity on defect texturing mechanisms. The particle- defect texture interaction in these systems is also discussed. Section 5 presents the conclusions of this work on defect texture and lattices in polygonal arrangements of cylindrical inclusions in cholesteric LC matrices.

## 2 Theory and Governing Equations

### 2.1 Description of Long-Range Orientational Order

As mentioned earlier, LC phases possess long range orientational order of their constituent molecules. This long range molecular order is described by an orientation distribution function, whose second moment is known as the quadrupolar symmetric and traceless tensor order parameter $\mathbf{Q}^{10,21}$:

$$\mathbf{Q} = S\left(\mathbf{nn} - \frac{\mathbf{I}}{3}\right) + \frac{P}{3}(\mathbf{mm} - \mathbf{ll}) = \mu_n \mathbf{nn} + \mu_m \mathbf{mm} + \mu_l \mathbf{ll} \qquad (2a)$$

where the following restrictions apply:

$$\mathbf{Q} = \mathbf{Q}^T \qquad (2b)$$

$$\text{tr}(\mathbf{Q}) = 0 \qquad (2c)$$

$$-\frac{1}{2} \leq S = 3(\mathbf{nn} : \mathbf{Q})/2 \leq 1, \qquad (2d)$$

$$-\frac{3}{2} \leq P = 3(\mathbf{mm} - \mathbf{ll}):\mathbf{Q}/2 \leq \frac{3}{2}, \qquad (2e)$$

$$\mathbf{n}.\mathbf{n} = \mathbf{m}.\mathbf{m} = \mathbf{l}.\mathbf{l} \qquad (2f)$$

The molecular orientation is defined completely by the orthogonal director triad (**n**, **m**, **l**). The measure of molecular alignment is defined by the uniaxial (S) and biaxial (P) scalar order parameters. The quadrupolar symmetry of the **Q** tensor retains the head-tail invariance *i.e.*, **n**(z) and −**n**(z) are equal. The cholesteric phase being periodic, the Q tensor satisfies the following condition: $\mathbf{Q}(z) = \mathbf{Q}(z + \mathbf{p}_o)$. The cholesteric phase is inherently biaxial (S ≠ 0, P ≠ 0) and the three eigenvalues $\mu_n$, $\mu_m$, $\mu_l$ are non-zero and distinct.

*2.2 Landau- de Gennes Theory for LC Materials*

The Landau- de Gennes (LdG) theory expresses the free energy density of the liquid-crystalline material in terms of homogeneous and gradient elastic contributions. The homogeneous (gradient) elastic effects are represented by power series of scalar invariants of tensor parameter **Q** (gradient of tensor order parameter ∇**Q**). In the absence of an external field, the total free energy density of the mesophase (f) can be given in the following dimensionless form[10,14,21,23]:

$$f = f_{iso} + f_{sr} + f_{lr} \qquad 3(a)$$

$$f_{sr} = \left\{ \frac{1}{2}\left(1 - \frac{U}{3}\right) tr\mathbf{Q}^2 - \frac{U}{3} tr\mathbf{Q}^3 + \frac{U}{4}\left(tr\mathbf{Q}^2\right)^2 \right\}, \quad\quad 3(b)$$

$$f_{lr} = \frac{1}{2}\left\{ \left[\left(\frac{\xi}{h_0}\right)(\nabla \times \mathbf{Q}) + 4\pi\left(\frac{\xi}{p_0}\right)\mathbf{Q}\right]^2 + \upsilon\left[\left(\frac{\xi}{h_0}\right)(\nabla \cdot \mathbf{Q})\right]^2 \right\} \quad\quad 3(c)$$

where $f_{iso}$ is the free energy density of the isotropic state which depends on conventional thermodynamic parameters, such as temperature, pressure, and concentration. $f_{sr}$ and $f_{lr}$ are, respectively, the short- and long-range contributions to the total free energy density f. The dimensionless parameter U is a thermodynamic potential proportional to the dimensionless concentration of anisotropic molecules which drives the isotropic- chiral-nematic phase transition; U is related to the concentration C by the relation U = 3C/C*, where C* is the concentration of the mesogens at the phase transition. The parameter $\xi$ is a coherence/internal length that gives the distance over which variations of long-range orientational order can occur. This length represents the order of magnitude for the size of the core of a disclination defect. The parameter $h_0$ is a macroscopic length scale external/geometric length that gives the size of the domain. The parameter $p_0$ is the pitch of the chiral-nematic liquid-crystalline material, the distance (along the director) over which the mesogens undergo a full 360° twist. It is essential to recognize that this model is therefore of a mesoscopic nature since it includes a molecular length scale ($\xi$) and a macroscopic ($h_0$) length scale. The remaining parameter $\upsilon$ represents a measure of the elastic anisotropy of the material. This parameter is constrained to be greater than 1/2 to ensure the positivity and thermodynamic stability of the long-range contribution to the free energy.

The nematodynamics equation used to describe the time evolution of the tensor order parameter **Q** is given by:

$$-\gamma\frac{\partial \mathbf{Q}}{\partial t} = \frac{\delta f}{\delta \mathbf{Q}} = \left[\frac{\partial f}{\partial \mathbf{Q}} - \nabla \cdot \frac{\partial f}{\partial \nabla \mathbf{Q}}\right]^{[s]} \quad\quad (4)$$

where the superscript [s] denotes symmetric and traceless tensors, and where $\gamma$ is a rotational viscosity[25]. Substituting eqn.(3) into eqn.(4) yields the equation for spatio-temporal evolution of tensor order parameter **Q**(**x**,t).

*2.3 Characterization Methods: Lubensky- de Gennes Coarse Grained Elastic Theory and layer deformation modes*

To guide the exploration of the model to significant biological ranges and to characterize the numerical results, we use the coarse grained model of Lubensky- de Gennes[10]. At scales larger than the equilibrium pitch $p_o$, cholesteric LC phases are layered structures characterized in terms of layer bending and dilation using a coarse grained elastic theory. According to the Lubensky- de Gennes theory, the dimensionless free energy density due to the layer displacement "u" reads[10]:

$$f = \left[ \underbrace{\frac{1}{2}K_1\left(\frac{\partial^2 u}{\partial x^2}\right)^2}_{\text{Bending}} + \underbrace{\frac{1}{2}B\left(\frac{\partial u}{\partial z}\right)^2}_{\text{Dilation}} \right] \qquad \text{5(a)}$$

$$\text{bending modulus } K_1 = \frac{27}{16}S^2\left(\frac{\xi}{h_0}\right)^2(1+\upsilon) \qquad \text{5(b)}$$

$$\text{dilation modulus } B = 9S^2(2\pi)^2\left(\frac{\xi}{p_0}\right)^2 \qquad \text{5(c)}$$

$$\text{penetration length } \lambda_1 = \sqrt{\frac{K_1}{B}} = 0.069\left(\frac{p_0}{h_0}\right)\sqrt{(1+\upsilon)} \qquad \text{5(d)}$$

The penetration length $\lambda_1$ defines the dilation strain required to produce bending in a set of parallel layers and it is smaller than the pitch $p_o$. The dimensionless bending ($K_1$) and dilation (B) parameters of this model are expressed in terms of the scalar order parameter (S), the length-scale ratios $\xi/h_0$, $\xi/p_0$, and the elastic anisotropy ($\upsilon$) of the original LdG model (eqn. (3)). Hence changes in the coarse grained parameters ($K_1$, B) that control the layer geometry can be predicted by changes in the primitive LdG parameters. As per eqns. (3,4), the defect textures given by **Q**(**x**,t) are modulated by the length scale ratios $(\xi/h_0), (\xi/p_0)$, the elastic anisotropy $\upsilon$ and the thermodynamic potential U. In this study, the parametric space has been limited to $(\xi/p_0)$ and $\upsilon$. Increasing elastic anisotropy decreases the dimensionless dilation/bending ratio:

$$\frac{B}{K_1} = 211.27 \frac{\left(\frac{h_0}{\xi}\right)^2 \left(\frac{\xi}{p_0}\right)^2}{(1+\upsilon)} \qquad (6)$$

Since a high value of $B/K_1$ (small $\lambda_1$) promotes defect nucleation, large anisotropy, large pitches, and smaller samples will tend to decrease defect proliferation in strongly twisted cholesterics. In terms of penetration length $\lambda_1$:

$$\lambda_1 = \sqrt{\frac{K_1}{B}} = 0.069 \left(\frac{1}{\xi/p_0}\right)\left(\frac{\xi}{h_0}\right)\sqrt{(1+\upsilon)} \qquad (7)$$

we see that large elastic anisotropy $\upsilon$, large pitch $p_o$, and smaller samples $h_o$ increase $\lambda_1$ which decreases defect content in strongly twisted cholesterics. The elastic anisotropy ratio in our model, relates to the well-known Frank's bending and twist elastic constants as: $\upsilon = \left[2(K_{33}/K_{22}) - 1\right]$ and assumes equal bend and splay deformations *i.e.*, $K_{11} = K_{33}$. For monomeric mesogens far from phase transitions, the elastic anisotropy is modest ($\upsilon \approx$ 1), but for polymeric rods it is large ($\upsilon \gg 1$), changing the relative energetic costs of bending/dilation and the defect density in the observed textures[13,26,27].

## 3 Computational Modeling

To understand the defect texturing during directed chiral self-assembly in cholesteric mesophases with embedded cylindrical inclusions, we simulate the time evolution of the tensor order parameter **Q** in 2D domains with a network of inclusions that forms: (a) an equilateral triangle (N=3), (b) a square (N=4), and (c) a regular hexagon (N=6), when joining their centers. To minimize the computational time, a single unit cell of finite size with periodic boundary condition is used to represent the 2D infinite periodic structure. As constructing an infinite periodic structure with only regular pentagons is not feasible, the analysis has been limited to triangular, square and hexagonal inclusion networks. All the length variables of the model are scaled with the macroscopic length scale of the model $h_0$. The dimensionless height and width of the domains with triangular, square and hexagonal inclusion networks are 0.866×1, 1×1 and 1.732×1.5 respectively. This geometry inherently generates anisotropic confinement in the domain of self assembly. The LC domain confined in a triangular network of inclusions is strongly confined, while that confined in an hexagonal network of inclusions is weakly confined. This aspect of the model arising from the configuration of particle arrangement is essential in understanding the effect of confinement on defect textures and lattices observed in our simulations. The dimensionless radius of the inclusions (R) is 0.15 and the dimensionless separation distance between the particles measured along the line connecting their centers ($r_0$) is 0.2 and are constant for all the three domains under investigation in section 4.1-4.4. In section 4.5, to study the effect of inclusion size on defect textures, the dimensionless radius of the inclusions (R) are varied from 0.125-0.175 and the dimensionless separation distance between the particles measured along the line connecting their centers ($r_0$) is 0.25- 0.15. At the surface of the particles represented by circles, the director exhibits a strong planar anchoring. These surfaces act as constraining layers that direct the LC phase ordering. It is also important to note that, although the simulation domain is two dimensional (2D), **Q** conserves its five degrees of freedom. Since the tensor order parameter has five independent components (symmetric and traceless), five coupled time-dependent nonlinear partial differential equations (eqn. 4) need to be solved simultaneously. Initially the LC material is taken to be in a stable isotropic state. This disordered state is imposed by using $\mathbf{Q}_0 = \mathbf{Q}(t = 0) = 0$. The governing equations are then solved using the Finite Element Method software package COMSOL, using adaptive finite elements and biquadratic basis functions with the generalized minimal residual (GMRES) solver method and a time-dependent backward Euler method. Convergence, mesh-independence, and stability are confirmed for each simulation result.

## 4  Results and discussion
### 4.1  Defect Textures and Charge Balance

This section presents simulation results of defect textures while varying the cholesteric pitch ($p_0$) and elastic anisotropy ($\upsilon$), while keeping the length scale ratio $\xi/h_0$ and themodynamic potential U constant, thus changing $K_1$, B and $\lambda_1$ (see eqn. 5(b-d)). The cholesteric layering is visualized by plotting the out-of-plane component $|n_z|$ of the director field (see eqn.(1)). Surface plots of the scalar order parameter S and z-component of the director $|n_z|$ are used to identify defect cores (fig. 3). All the results presented in this work are steady state solutions. The type, number, strength, and charge of all the disclination lines and dislocations in the LC domain confined within a polygon of N sides obtained by joining the centers of N cylindrical inclusions for N = 3,4,6 and the type of LC phase as predicted by eqns.(3,4) as a function of length scale ratio $\xi/p_0$ and elastic anisotropy $\upsilon$, for a fixed inclusion size and separation distance are shown in Table 1. The length scale ratio $\xi/h_0$ is kept constant for all the simulations at 0.0025. The coherence length ($\xi$) is of the order of disclination core size and is about 10nm, the chosen value of length scale ratio $\xi/h_0$ fixes the inclusion diameter around 1.2µm and is fixed for all the simulations and the minimum separation distance between two inclusions around 0.8 µm. In an attempt to study the effect of cholesteric pitch on the defect texture and LC phases arising in our simulations, the length scale ratio $\xi/p_0$ is varied from 0.05 to 0.001, which corresponds to variation of cholesteric pitch from 0.2 to 10µm. The effect of the elastic anisotropy has been investigated by computing the defect textures and phases for $\upsilon=1$ (corresponds to equal bend, splay and twist elastic modulii) and $\upsilon=21$ (corresponds to bend and splay modulii equal to 11 times the twist elastic modulus).

In Table 1, the first column defines the polygonal arrangement *N*, the second the length scale ratio $\upsilon$, the third the elastic anisotropy constant $\xi/p_0$, the fourth the number, charge and the type of bulk defects, the fifth the type of LC phase. It is evident that the presence of a polygonal network with N circular inclusions results in defect textures of a net charge of –(N–2)/2 per unit polygonal cell, as predicted by Zimmer's rule[5,17]. This prediction is in agreement with the observed $\lambda^{-1}$ disclination in the domain bound between four cells in the cell wall of a walnut[2] (Fig. 1) and also the theoretical free energy computation of a hypothetical isotropic region bound between three double twist cylinders resolving to be a -1/2 disclination[10].

The LC phases observed in the simulation are depicted in terms of surface plots of z-component of the director $|n_z|$ and the scalar order parameter S in fig. 3. Fig. 3(a) and 3(b) shows a fingerprint texture of cholesteric phase with characteristic "stripes" and dislocations and disclinations, filled with cylindrical particles[28]. Fig. 3(c) and 3(d) shows cholesteric phase with only disclinations. Fig. 3(e) and 3(f) shows 2D cholesteric hexagonal blue phase with singular disclination lattice and all double twist cylinders replaced by cylindrical particles. Fig. 3(g) and 3(h) shows 2D cholesteric hexagonal blue phase with non-singular disclination lattice and some of the double twist cylinders replaced by cylindrical particles. Fig. 3(i) and 3(j) shows 2D cholesteric tetragonal blue phase with double twist cylinders replaced by cylindrical particles. Fig. 3(k) and 3(l) shows weakly twisted filled cholesterics with singular nematic disclinations. Fig. 3(m) and 3(n) shows cholesteric phase with high strength disclinations filled with cylindrical particles. Singular disclinations are identified from the scalar order parameter surface plots and non-singular disclinations are identified from |$n_z$| surface plots. The mechanism behind formation of these LC phases and defects textures are detailed in section 4.2-4.4 using Lubensky- de Gennes coarse grained elastic theory detailed in Section 2.3

*4.2 Effect of the cholesteric pitch on the defect textures and LC phases*

To gain insights into the role of the cholesteric pitch ($p_0$) and confinement on the type of the LC phase and defect texture arising from a chiral self assembly due to cylindrical inclusions, in this section the length scale ratio $\xi/p_0$ is decreased from 0.05 to 0.005. This variation corresponds to increase in cholesteric pitch from 0.2 to 2μm. The LC phases are computed for three domains with different polygonal arrangement of inclusions, each representing different degrees of confinement. The length scale ratio $\xi/h_0$, thermodynamic potential U and elastic anisotropy constant $\upsilon$ are kept constant at 0.0025, 6 and 1. This corresponds to a variation of the dimensionless penetration length ($\lambda$) from $4.879\times10^{-3}$ to 0.04879. Since the penetration length for all the simulations in this section are less than 1, it is easier to deform the layers through bending than to compress/dilate the layers. Hence for all the textures observed in this study, the layer thickness is approximately equal to $p_0/2$. The results and discussion in sections 4.2 and 4.3 are limited to monomeric mesogens far from phase transitions.

Since the LC material of layer periodicity $p_0/2$ is confined between two inclusions apart by a separation distance $r_0$, the transitions between different LC phases and their corresponding elastic energies can be described as a function of the length scale ratio $r_0/(p_0/2)$. The elastic energy of the LC phases is characterized by the excess free energy per particle, the difference between the total free energy of the LC domain with an inclusion and the free energy of the same volume of defect free cholesteric planar monodomain.

For a 2D domain, the excess free energy $f_e$ [15] is:

$$f_e = \frac{\int_A (f_g + f_h) dA - f_m A}{N_a} \tag{8}$$

where $f_m$ is the free energy of the defect free monodomain, given by[23]

$$f_m = \left(-\frac{1}{9}S^2 - \frac{1}{27}P^2\right)U + \frac{1}{3}S^2 + \frac{1}{9}P^2 - \frac{2}{27}US(S^2 - P^2) + \frac{1}{81}U(3S^2 + P^2)^2 + \frac{4}{3}\pi^2\left(\frac{\xi}{p_0}\right)^2 (S-P)^2 \tag{9}$$

and where A is the area of an unit cell of the domain; $N_a$ is the number of particles in a unit cell of a domain (for example, in Fig. 3(g) unit cell of a domain with hexagonal particle network has 8 particles). We note that the excess free energy density $f_e$ is a measure of the total energy increase per particle above that of the defect free planar monodomain energy $f_m A$, due to layer distortions (bending and dilation) and defect nucleation.

Fig. 4 shows the dimensionless excess free energy per cylindrical inclusion as a function of $r_0/(p_0/2)$, for three different polygonal arrangement of cylindrical inclusions (N=3,4,6). For a given particle configuration, when $2r_0/p_0 \gg 1$, the impingement of layers growing from different particles and the nucleation of defects in the domain of maximum stress concentration induced by confinement (at the intersection of diagonals of the polygon) results in a phase rich in dislocations and disclinations. The resulting phase has stripes typical of fingerprint cholesteric textures filled with cylindrical inclusions (fig. 3(a) and (b)). The excess free energy per particle associated with these phases is high due to the presence of numerous singular disclination cores of high elastic energy cost. For $2r_0/p_0 \geq 1$, the resulting phase has only disclinations. These disclinations may arise due to impingement of layers and/or induced by stress concentration under confinement, resulting in particle filled cholesteric phases with only disclinations or particle filled 2D blue phases with singular disclination lattice (fig. 3(c) and (d)). The excess free energy per particle computed for these domains are lesser than that of fingerprint textures, owing to relatively lesser number of singular disclination cores. When $2r_0/p_0=0.8$, the resulting phase has only non-singular disclinations induced by stress concentration under confinement, resulting in particle filled cholesteric blue phases with non-singular disclination lattice (fig. 3(g)-(j)). These phases represent lowest elastic energy, since all the defects nucleated have non-singular disclination cores. Hypothetically, when planar cholesteric layers of pitch $p_0$ is confined between two flat surfaces separated by a distance $r_0$, the length scale ratio $2r_0/p_0=1$ would have resulted in the minimum excess free energy as one planar cholesteric layer of thickness half a pitch ($p_0/2$) would have occupied the distance between two adjacent particles $r_0$ without deformation of layers or any orientational incompatibility between director fields at the line of impingement. In our simulations, the minimum excess free energy per particle is observed at $2r_0/p_0=0.8$ owing to the relative ease of compression/ dilation of the layers as the penetration length approaches 1 and as the curvature of the constraining surface are propagated into the bulk through surface anchoring and gradient elasticity. As the length scale ratio $2r_0/p_0$ tends to 0, the weakly twisted cholesteric phases[29] with thin singular line disclinations distinctive of nematic LC phases are stabilized (fig. 3(k) and (l)). In our simulations, these defect cores are singular ($s=-1/2$ and $s=+1/2$) resulting in higher elastic energy costs. Hence, for a given polygonal arrangement in a chiral mesophase composed of monomeric mesogens, the particle separation distance to equilibrium cholesteric layer thickness ratio $2r_0/p0$ of 0.8 favours 2D blue phases.

*4.3 Effect of particle configuration on the elastic energy of the LC phases*

To identify the polygonal configuration that may lower the elastic energy, the excess free energy per particle for phases arising under different polygonal arrangements are compared in this section.

From Fig. 4, it is evident that 2D hexagonal blue phases partially filled with cylindrical inclusions arising in chiral self-assembly in hexagonal arrangement of inclusions (N=6) at low chirality ($2r_0/p0=0.8$) have the lowest elastic energy. For the range of cholesteric pitches explored in this study, phases with fourfold square symmetry observed under chiral self-assembly in square arrangement of particles have the highest elastic energy. This remark is in agreement with theoretical and experimental observations that at low chirality (large pitch), in the absence of external fields, the hexagonal blue phase appears as an intermediate phase during isotropic to cholesteric transition rather than tetragonal blue phases[30]. The hexagonal blue phase with cylindrical inclusions occupying all non-singular $\lambda^{+1}$ disclinations arising under triangular arrangement of particles has slightly higher elastic energy than partially filled hexagonal blue phase arising under hexagonal particle arrangement (fig. 3(g) and (h)), owing to higher defect density per particles per unit area.

At lower values of $2r_0/p_0$ ($2r_0/p_0 \leq 4$), triangular arrangement of particles that impose stronger confinement and higher number of particles per unit area of the domain has higher elastic energy than hexagonal arrangement of particles that impose weaker confinement. However, the decrease of difference in elastic energy at the onset of fingerprint textures as the pitch is increased suggests the elastic energy of a fingerprint texture formed under hexagonal arrangement of particles might exceed that of square arrangement at a higher chirality. This predicted increase in elastic energy of fingerprint textures formed under hexagonal arrangement can be seen in fig. 4, comparing the excess free energy per particle for phases arising under triangular and hexagonal particle arrangement. The elastic energy of a fingerprint texture formed under hexagonal arrangement of particles surpasses that of triangular arrangement at $2r_0/p_0=6.7$. This can be attributed to competition between core energies of defects nucleated through multiparticle interaction at the domain of stress concentration with that of the defects nucleated through particle-particle interaction due to impingement of layers growing from two particles. At low chirality, for a given symmetry of the LC phase, the particle with weaker (stronger) confinement results in a phase with lower (higher) elastic energy. As the chirality of the LC phase is increased, the defect nucleation through Kibble's mechanism, where layers growing from different particles impinge, resulting orientational incompatibilities that are relived through nucleation of dislocations and disclination, is initiated. As the chirality is increased further, the number of layers impinging and hence the defects nucleated are increased proportionally, while the number of defects due to confinement effects does not vary considerably. This leads to a regime where the elastic energy of a LC phase is proportional to the number of particles that forms the polygonal network. Hence at higher chirality, a LC phase formed in a hexagonal particle network has higher elastic energy than one formed in square and triangular particle networks. In these phases, the arrangement of defects due to layer impingement exhibits the same symmetry as that of the particle arrangement.

*4.4 Effect of elastic anisotropy on the defect textures and LC phases observed*

Elastic anisotropy is a crucial material property in defect texture studies as it plays a vital role in defect dynamics and preferred modes of layer distortion and hence the nature of defects that are stabilized. As the elastic energy variation with cholesteric pitch is monotonous and the phase transitions are identical for all three embedded polygonal network of inclusions (Fig. 4), chiral self assembly with square arrangement of particles in the domain (N=4) is used as a model configuration for studying the effect of elastic anisotropy in this section. The length scale ratio $\xi/p_0$ is varied from 0.05 to 0.005, and two values of elastic anisotropy constant $\upsilon=1$ corresponding to monomeric mesogens far from phase transitions and $\upsilon=21$ corresponding to semiflexible polymeric mesogens relevant to biological chiral self assembly. The length scale ratio $\xi/h_0$ and thermodynamic potential U are kept constant at 0.0025 and 6 respectively. The surface plots of z-component of the director $|n_z|$ for LC phases observed at high elastic anisotropy in the simulation are depicted in fig. 5.

Fig. 6 shows the dimensionless excess free energy per cylindrical inclusion as a function of $r_0/(p_0/2)$, for two values of elastic anisotropy ($\upsilon=1, 21$) for N=4. For a given cholesteric pitch, the increase in elastic anisotropy increases the penetration length for polymeric mesogens by 3.317 times. This makes the layers in polymeric chiral mesophases relatively easier to dilate/compress and difficult to bend, compared to monomeric mesogens. As a result, disclinations with high strength that are unstable at $\upsilon=1$ are stabilized at higher elastic anisotropy $\upsilon=21$ at lower chirality. Stabilization of $\lambda^{-2}$ disclination with six-fold hexagonal symmetry in domain confined in an hexagonal particle arrangement at low chirality and high anisotropy (polymeric mesogens), indicates particle symmetry controlled defect nucleation that is only observed at high chirality for monomeric mesogens (fig. 3 (m) and (n)).

At high chirality and high elastic anisotropy (polymeric mesogens), fingerprint textures with fewer dislocations/ dislocations arise. The observed defect textures have the following characteristics: (a) the symmetry of the particle arrangement is not transferred to the defect arrangement and the texture becomes more random (compare fig. 3(a) and 5(b)), (b) non-singular disclinations of strength +1 ($\lambda^{+1}$) that involve bending of layers are now energetically costly and are avoided, and (c) the layers tend to avoid bending and are locally planarly aligned through: (i) the nucleation of defects closer to the particle surface and (ii) the nucleation of quadrupolar defects such as Lehmann clusters, that are known to arise in nematic[31], cholesterics[32] and smectic LCs[33]. This result is in good agreement with helicoidal textures formed from cellulose microfibrils through self assembly in the cell wall of a prune with a triangular network of pit canals (P) embedded in cholesteric mesophase[2] (Fig. 1(b)). The observed textures also follow experimental observations in three helical biological polymeric mesophases, namely PBLG (a polypeptide), DNA (a polynucleotide) and xanthan (a polysaccharide)[34]. Despite their different chemical nature, the phases given by these biopolymers are devoid of solitary $\tau^{-1/2}$ disclinations. If present at all, these singular disclinations are observed only as $\tau^{-1/2}\lambda^{+1/2}$ disclination pairs. The results of our simulations at high elastic anisotropy corresponding to LC phases of semi-flexible polymeric mesogens (Table 1, Fig. 5) are consistent with this experimental observation.

For semiflexible polymeric mesogens (high elastic anisotropy), defect textures with fewer disclinations/ dislocations arise but due to layer distortions (bending and dilation) we find a higher elastic energy than monomeric mesogens (fig. 5).

*4.5 Defect lattice – particle interaction in chiral self-assembly systems*

To understand the interaction of particles with the LC phases and defect textures observed, we simulate the 2D blue phases without embedded particles. The effect of particle size is also investigated. The 2D blue phases with only non-singular defect cores are simulated by dictating mesogenic orientation along the third direction at points representing the center of the particles. Depending on the arrangement of the points, tetragonal 2D blue phase with superimposed $\lambda^{+1}$ and $\lambda^{-1}$ disclination lattices with fourfold square symmetry or hexagonal 2D blue phase with $\lambda^{+1}$ disclination lattice with sixfold symmetry superimposed by $\lambda^{-1/2}$ disclination lattice with threefold symmetry are formed. By comparing the filled 2D blue phases observed in our simulations, we conclude that the particles occupy $\lambda^{+1}$ disclinations cores thus reducing the free energy cost associated with these disclinations and has a stabilizing effect on the LC phases[4,35]. Similar phenomenon has been observed in nanoparticles in LC phases[32], and spherical colloidal particles in cholesteric phases[35].

Under strong anchoring at the particle interface, the director field deviates from that of a 2D blue phase due to distortions induced by the inclusion of the particles. These distortions give rise to an anchoring penalty which is a function of particle size $(R)^{32}$. Of the three particle sizes explored in this study (R=0.125, 0.15 and 0.175, $\xi/p_0$=0.005, $\upsilon$=1), particles of dimensionless radius 0.15 exhibit least deviation from 2D blue phase director field and hence the minimum anchoring penalty. All the three particle sizes explored result in single $\lambda^{-1}$ disclination. At values of R<0.125, cholesteric phases with disclinations emerge while for R>0.175, the LC phase is a weakly twisted cholesteric. This anchoring penalty can be eliminated by using weak anchoring at the particle surface[35]. However, for micron size particles investigated in this work, the free energy cost associated with anchoring penalty is insignificant compared to the energy reduction due to the particles replacing $\lambda^{+1}$ disclinations.

## 5 Conclusions

A systematic analysis of defect textures in polygonal arrangement of micron-sized cylindrial inclusions in a cholesteric LC matrix has been performed using the Landau de Gennes model for chiral self assembly. The effect of varying cholesteric pitch, particle configuration, elastic anisotropy and particle size on defect lattices, textures and LC phases stabilized have been investigated in the ideal case of arrangements of particles with perfect order. The presence of a polygonal network with N circular inclusions results in defect textures of a net charge of –(N–2)/2 per unit polygonal cell, as predicted by Zimmer's rule (Table 1). This prediction is in agreement with the observed $\lambda^{-1}$ disclination in the domain bound between four cells in the cell wall of a walnut[2] (Fig. 1) and theoretical free energy computation of a hypothetical isotropic region bound between three double twist cylinders resolving to be a -1/2 disclination[10]. As the chirality is increased, the LC matrix shows the following series of transitions: weakly twisted cholesterics, 2D blue phases with non-singular/ singular defect lattices, cholesteric phases with only disclinations, fingerprint cholesteric textures with disclinations and dislocations (Fig. 3). The elastic energy of the domain represented in terms of excess free energy per particle monotonically increases with increasing chirality (Fig. 4) for strongly twisted cholesterics. Depending on the particle configuration, the LC matrix phase might be strongly/ weakly confined. In monomeric mesogens, for a given symmetry of the LC phase, the particle with weaker (stronger) confinement results in a phase with lower (higher) elastic energy at low chirality. Hexagonal symmetry of the LC phases is energetically favoured in comparisons to tetragonal symmetry. At high chirality, the elastic energy of a LC phase is proportional to the number of particles that form the polygonal network. The arrangement of defects due to layer impingement exhibits the same symmetry as that of the particle arrangement. Hexagonal (triangular) particle arrangement results in low elastic energy at low (high) chirality. For semiflexible polymeric mesogens (high elastic anisotropy) the observed defect textures have the following characteristic. (I) At high chirality: (a) The symmetry of particle arrangement is not transferred to the defect arrangement and the texture becomes more random, (b) non-singular disclinations of strength +1 ($\lambda^{+1}$) that involve bending of layers are now energetically costly and are avoided, and (c) the layers tend to avoid bending and are locally planarly aligned through: (i) the nucleation of defects closer to the particle surface and (ii) the nucleation of quadrupolar defects such as Lehmann clusters; (II) At low chirality: (d) high strength disclination lines possessing the same symmetry as the particle arrangement are stabilised. For semiflexible polymeric mesogens (high elastic anisotropy), defect textures with fewer disclinations/ dislocations arise but due to layer distortions (bending and dilation) we find a higher elastic energy than monomeric mesogens (fig. 5).

The defects arising in the simulations and the texture rules established are in agreement with experimental observations in cellulosic liquid crystal analogues such as plant cell wall and helical biological polymeric mesophases made of DNA, PBLG and xanthan. The above-mentioned findings are compactly summarized in the phase diagram (fig. 7), that shows the different LC phases and textures as a function of chirality and elastic anisotropy. Comparing the filled 2D blue phases observed in our simulations with 2D blue phases without any inclusions, we conclude that the particles occupy $\lambda^{+1}$ disclinations cores thus reducing the free energy cost associated with these disclinations. Thus the inclusion of particles has a stabilizing effect on the LC phases. Under strong anchoring at the particle interface, the director field deviates from that of a 2D blue phase due to distortions induced by the inclusion of the particles. These distortions give rise to anchoring penalty which is a function of particle size. However, for micron size particles investigated in this work, the free energy cost associated with anchoring penalty is insignificant comparing to that reduced by particles replacing $\lambda^{+1}$ disclinations. The findings provide a comprehensive set of trends and mechanisms that contribute to the evolving understanding of biological plywoods and serve as a platform for biomimetic applications.

## Acknowledgements

This work was supported by the U.S. Office of Basic Energy Sciences, Department of Energy, grant DE-SC0001412. Y.M was supported by grants from the Natural Science and Engineering Research Council of Canada and Fonds de recherche du Québec - Nature et technologies.

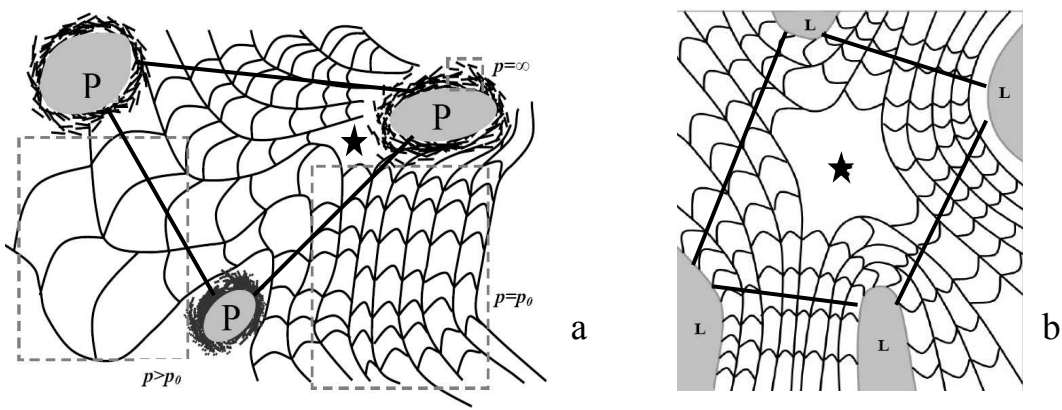

Fig. 1 Polygonal networks in some plant cell walls. (a) A triangular network of pit canals (P) embedded in cholesteric mesophase (represented by arced patterns) in cell wall of a prune[2] (b) LC phase with a rectangular network formed by four cell lumens (L) embedded in cell wall of a walnut[2]. In both the figs, the defect nucleated due to presence of inclusions is represented by '★' [Adapted from A. D. Rey, *Soft Matter*, 2010, **6**, 3402-3429].

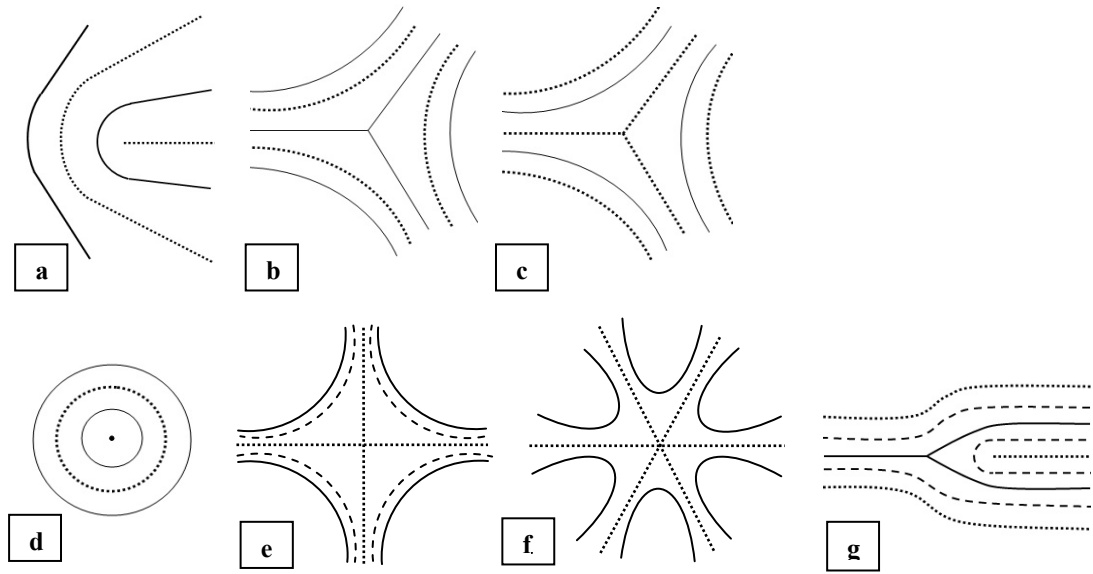

Fig. 2 Director configuration of some of the classical disclinations and disclination pairs arising in our simulations. (a) Non-singular $\lambda^{+1/2}$ disclination; (b) singular $\tau^{-1/2}$ disclination; ; (c) non-singular $\lambda^{-1/2}$ disclination; (d) non-singular $\lambda^{+1}$ disclination; (e) non-singular $\lambda^{-1}$ disclination; (f) high strength non-singular $\lambda^{-2}$ disclination; (g) $\tau^{-1/2}\,\lambda^{+1/2}$ disclination pair. The dotted lines indicate the director is pointing out of the plane of the diagram, the solid lines indicate the director is in the plane of the diagram, and the dashed line indicates that the director is tilted at an angle to the plane of the diagram.

**Table 1.** Summary of number and types of defects as a function of elastic anisotropy $\upsilon$, length scale ratio $\xi/p_0$ and number of sides of the unit polygonal cell N. For all the simulations, length scale ratio $\xi/h_0=0.0025$ and thermodynamic potential U=6 (see Fig. 2 for orientation nz and scalar order S fields).

| N | $\upsilon$ | $\xi/p_0$ | Defects in a unit polygonal cell | Type of filled LC phases |
|---|---|---|---|---|
| 3 | 1 | 0.05 | $6\times\tau^{-1/2}\lambda^{+1/2} + 3\times\tau^{-1/2} + 1\times\lambda^{+1}$ | Cholesteric fingerprint texture |
|  |  | 0.025 | $3\times\tau^{-1/2} + 1\times\lambda^{+1}$ | Cholesterics with disclinations |
|  |  | 0.0125 | $3\times\tau^{-1/2} + 1\times\lambda^{+1}$ |  |
|  |  | 0.01 | $1\times\tau^{-1/2}$ | Hexagonal 2D blue phase with singular disclination lattice Fig. 2(e) and (f) |
|  |  | 0.005 | $1\times\lambda^{-1/2}$ | Hexagonal 2D blue phase with non-singular disclination lattice |
|  |  | 0.00125 | 1, s=-1/2 (singular) | Weakly twisted cholesterics |
| 4 | 1 | 0.05 | $24\times\tau^{-1/2}\lambda^{+1/2} + 4\times\tau^{-1/2} + 1\times\lambda^{+1}$ | Cholesteric fingerprint texture Fig. 2(a) and (b) |
|  |  | 0.025 | $8\times\tau^{-1/2}\lambda^{+1/2} + 6\times\tau^{-1/2} + 2\times\lambda^{+1}$ |  |
|  |  | 0.0125 | $6\times\tau^{-1/2} + 2\times\lambda^{+1}$ | Cholesterics with disclinations Fig. 2(c) and (d) |
|  |  | 0.01 | $4\times\tau^{-1/2} + 1\times\lambda^{+1}$ |  |
|  |  | 0.005 | $1\times\lambda^{-1}$ | Tetragonal 2D blue phase Fig. 2(i) and (j) |
|  |  | 0.0005 | $2\times$ s=-1/2 (singular) | Weakly twisted cholesterics |
|  | 21 | 0.05 | $4\times\lambda^{-1/2} + 2\times\lambda^{+1/2} + 6\times\tau^{-1/2}\lambda^{+1/2}$ | Cholesteric fingerprint texture fig. 5(b) |

|   |   | 0.0125 | $1\times\lambda^{-1}$ | Tetragonal 2D blue phase Fig. 5(a) |
|---|---|---|---|---|
|   |   | 0.005 | $1\times\lambda^{-1}$ |  |
| 6 | 1 | 0.05 | $36\times\tau^{-1/2}\lambda^{+1/2} + 8\times\tau^{-1/2} + 2\times\lambda^{+1}$ | Cholesteric fingerprint texture |
|   |   | 0.025 | $6\times\tau^{-1/2}\lambda^{+1/2} + 8\times\tau^{-1/2} + 2\times\lambda^{+1}$ |  |
|   |   | 0.0125 | $6\times\tau^{-1/2} + 1\times\lambda^{+1}$ | Partially filled cholesteric hexagonal blue phase with singular disclination lattice |
|   |   | 0.01 | $6\times\tau^{-1/2} + 1\times\lambda^{+1}$ |  |
|   |   | 0.005 | $6\times\lambda^{-1/2} + 1\times\lambda^{-1}$ | Partially filled cholesteric hexagonal blue phase with non-singular disclination lattice Fig. 2(g) and (h) |
|   |   | 0.0001 | $6\times s=-1/2 + 2\times s=+1/2$ (Singular) | Weakly twisted cholesterics Fig. 2(k) and (l) |
|   | 21 | 0.005 | $1\times\lambda^{-2}$ | Cholesterics with high strength disclinations Fig. 2(m) and (n) |

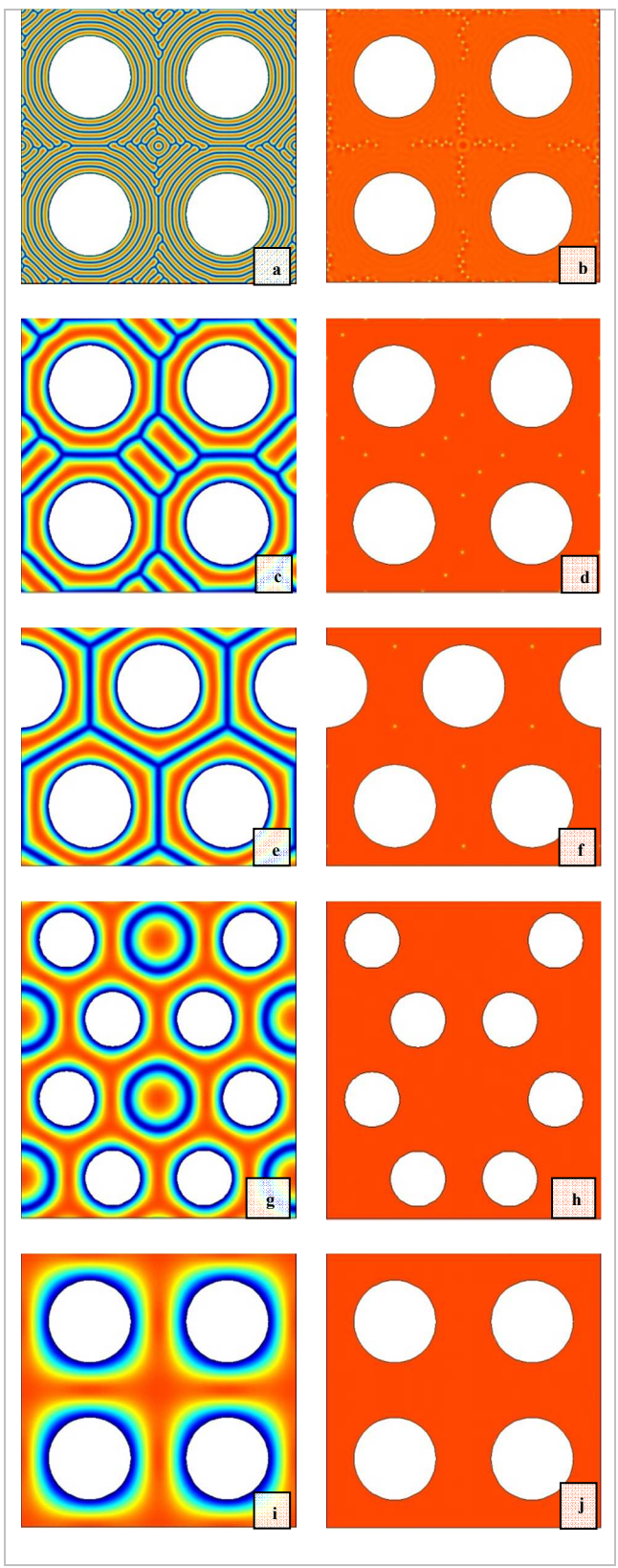

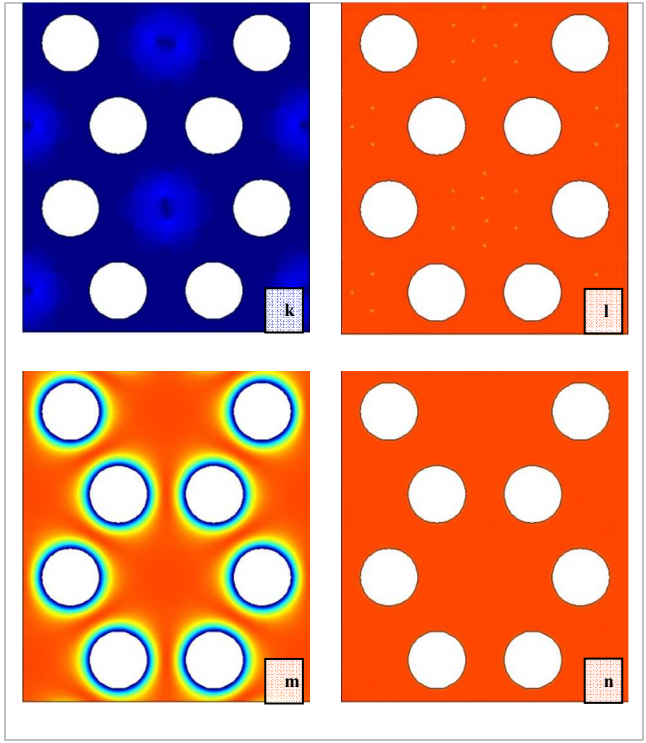

Fig. 3 Surface plots of z-component of the director $|n_z|$ (left) and the scalar order parameter S (right) of the observed LC phases. The fields go from blue to red as the out-of-plane component $|n_z|$ goes from 0 to ±1 and scalar order parameter S goes from zero to the equilibrium value. (a) and (b) Cholesteric fingerprint texture $\upsilon=1$, $\xi/p_0=0.05$, N=4; (c) and (d) Cholesterics with disclinations $\upsilon=1$, $\xi/p_0=0.0125$, N=4; (e) and (f) Hexagonal blue phase with singular disclinations $\upsilon=1$, $\xi/p_0=0.01$, N=3; (g) and (h) Hexagonal 2D blue phase with non-singular disclinations $\upsilon=1$, $\xi/p_0=0.005$, N=6; (i) and (j) Tetragonal 2D blue phase with non-singular disclinations $\upsilon=1$, $\xi/p_0=0.005$, N=4; (k) and (l) Weakly twisted cholesterics $\upsilon=1$, $\xi/p_0=0.0001$, N=6; (m) and (n) Cholesterics with high strength disclinations $\upsilon=21$, $\xi/p_0=0.005$, N=6. Singular disclinations are identified in the scalar order parameter surface plots (b,d,f,l).

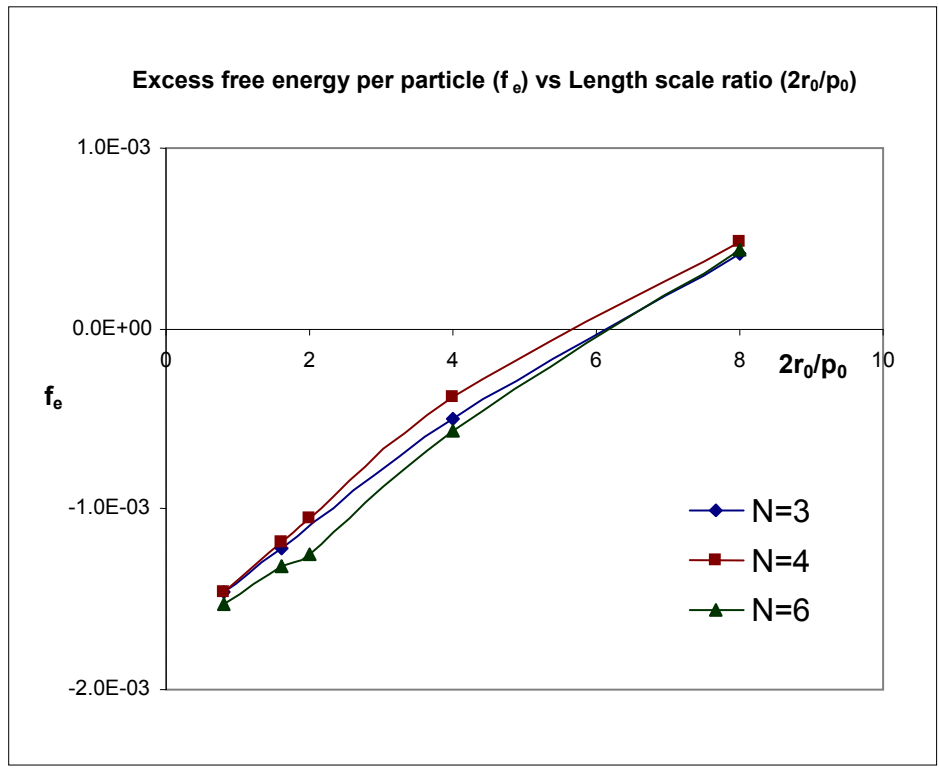

Fig. 4 Dimensionless excess free energy per cylindrical inclusion ($f_e$) as a function of length scale ratio $r_0/(p_0/2)$, for three different polygonal arrangements of cylindrical inclusions (N=3,4,6). As the length scale ratio ($2r_0/p_0$) is increased, the LC matrix shows the following series of transitions: weakly twisted cholesterics, 2D blue phases with non-singular/ singular defect lattices, cholesteric phases with only disclinations, fingerprint cholesteric textures with disclinations and dislocations. For all particle configurations, the excess free energy per particle, increases monotonously with increase in chirality.

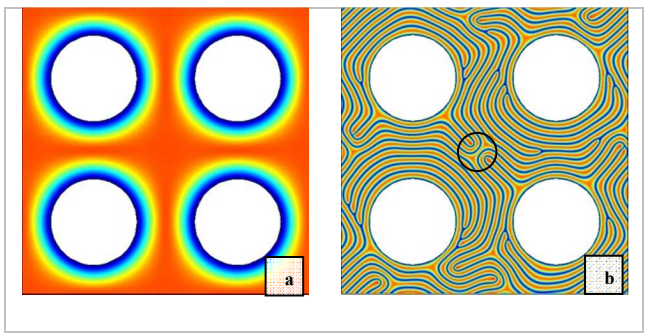

Fig. 5 Surface plots of the z-component of the director $|n_z|$ of the observed LC phases at high elastic anisotropy. The fields change from blue to red as the out-of-plane component $|n_z|$ varies from 0 to ±1. (a) Tetragonal 2D blue phase at high elastic anisotropy $\upsilon=21$, $\xi/p_0=0.0125$, N=4 (b) Cholesteric fingerprint texture at high elastic anisotropy $\upsilon=21$, $\xi/p_0=0.05$, N=4. The circled region encompasses Lehmann cluster, a quadrupolar defect interaction.

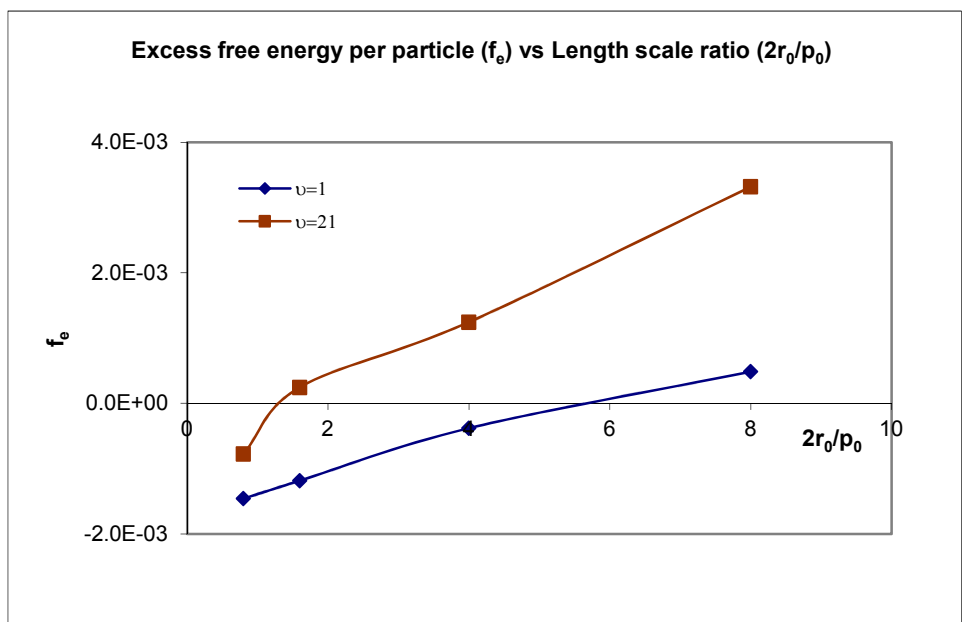

Fig. 6 Dimensionless excess free energy per cylindrical inclusion ($f_e$) as a function of length scale ratio $r_0/(p_0/2)$, for monomeric mesogens ($\upsilon=1$) and semi-flexible polymeric mesogens ($\upsilon=21$) for square arrangement of cylindrical inclusions (N=4). For the same configuration of cylindrical inclusions, increasing elastic anisotropy results in higher excess free energy per particle.

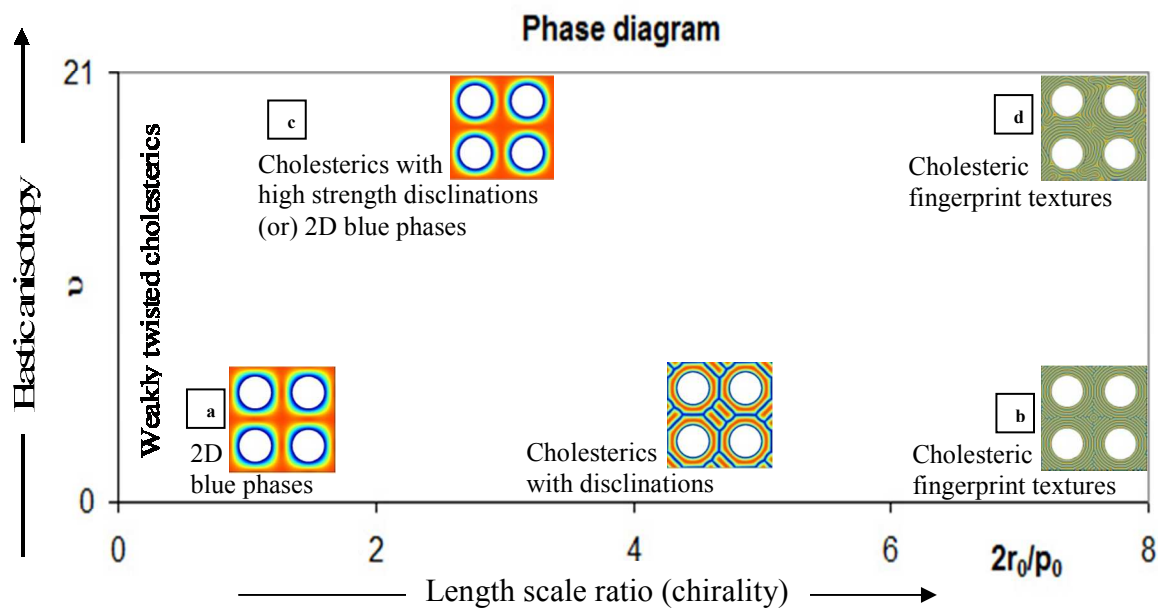

Fig. 7 Phase diagram for different LC phases as a function of length scale ratio $r_0/(p_0/2)$ and elastic anisotropy ($\upsilon$). (a) At low elastic anisotropy and low chirality, defects due to layer impingement are absent and the elastic energy of the LC phase depends solely on defects nucleated through confinement effects. (b) As the chirality is increased, disclinations and dislocations nucleated through layer impingement increase while the number of defects due to confinement effects does not vary considerably. This leads to a regime where the elastic energy of a LC phase is proportional to the number of particles that forms the polygonal network. (c) At high elastic anisotropy and low chirality, cholesterics with high strength disclination that has the symmetry of the particle arrangement (N=6) or 2D blue phases (N=3,4) are stabilized. (d) At high elastic anisotropy and high chirality, the resulting defect texture lacks the symmetry of the particle arrangement and quadrupolar defect interactions such as Lehmann cluster are observed.

TOC text:

A systematic computational and scaling analysis of defect textures in polygonal arrangement of cylindrical particles embedded in a cholesteric (Ch) liquid crystal matrix is performed using the Landau de Gennes model for chiral self assembly.

TOC diagram:

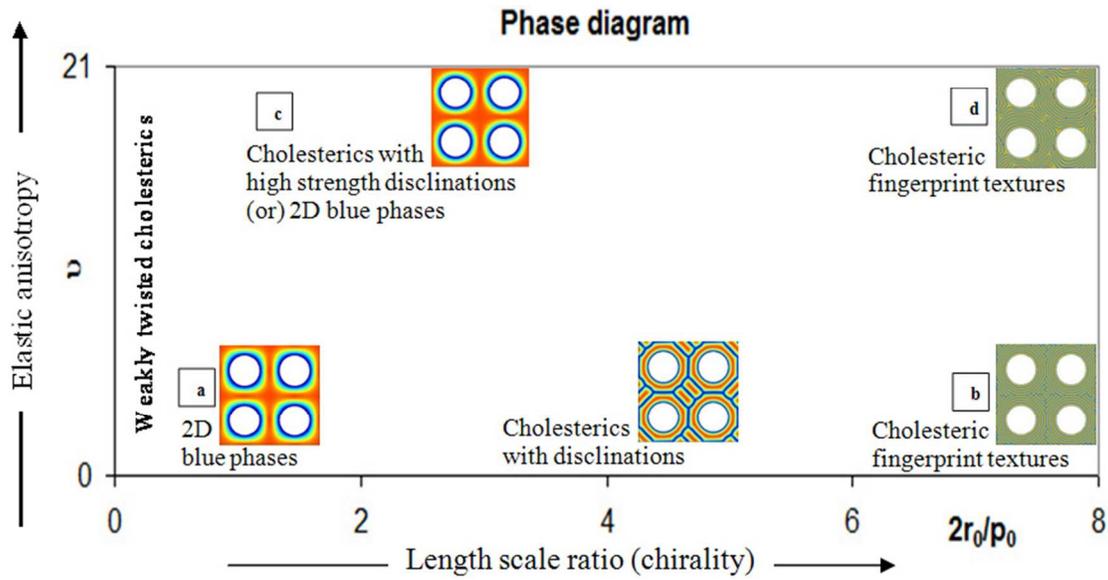

36